# Transient Foreshock Structures Upstream of Mars: Implications of the Small Martian Bow Shock

H. Madanian[1], N. Omidi[2], D. G. Sibeck[3], L. Andersson[1], R. Ramstad[1], S. Xu[4], J. R. Gruesbeck[3], S. J. Schwartz[1], R. A. Frahm[5], D. A. Brain[1], P. Kajdîc[6], F. G. Eparvier[1], D. L. Mitchell[4], S. M. Curry[4]

[1]Laboratory for Atmospheric and Space Physics, University of Colorado, Boulder, CO, USA

[2]Solana Scientific Inc., Solana Beach, CA, USA

[3]NASA Goddard Space Flight Center, Greenbelt, MD, USA

[4]Space Sciences Laboratory, University of California, Berkeley, CA, USA

[5]Southwest Research Institute, San Antonio, TX, USA

[6]Instituto de Geofísica, Universidad Nacional Autónoma de México, Mexico City, Mexico

Corresponding author: Hadi Madanian (hmadanian@gmail.com)

**Key Points:**

- Foreshock bubbles can form upstream of Mars.
- Slow field rotations can cause foreshock bubbles while reflected ions from the quasi-perpendicular bow shock contribute to their formation.
- Unique ion kinetic scale processes exist around foreshock structures at Mars due to the different interaction size scale.

**Abstract**

We characterize the nature of magnetic structures in the foreshock region of Mars associated with discontinuities in the solar wind. The structures form at the upstream edge of moving foreshocks caused by slow rotations in the interplanetary magnetic field (IMF). The solar wind plasma density and the IMF strength noticeably decrease inside the structures' core, and a compressional shock layer is present at their sunward side, making them consistent with foreshock bubbles (FBs). Ion populations responsible for these structures include backstreaming ions that only appear within the moving foreshock, and accelerated reflected ions from the quasi-perpendicular bow shock. Both ion populations accumulate near the upstream edge of the moving foreshock which facilitates FB formation. Reflected ions with hybrid trajectories that straddle between the quasi-perpendicular and quasi-parallel bow shocks during slow IMF rotations contribute to formation of foreshock transients.

**Plain Language Summary**

Planets in the solar system are continuously impacted by the solar wind, a plasma flow originating at the Sun and propagating through the interplanetary medium at high speeds. The solar wind also carries a magnetic field which at times contains twists or discontinuities. The discontinuities are associated with large scale electric currents that can have planar shapes. A planetary obstacle significantly modulate the solar wind plasma and the interaction of solar wind discontinuities with the modulated plasma upstream of the planet leads to formation of transient structures. Due to their relatively large size, these structures can significantly impact and destabilize plasma boundaries at lower altitudes closer to the surface. The results of this paper improve our understanding of solar wind interactions and formation of transient structures upstream of Mars.

**1 Introduction**

Plasma environments and the stability of plasma boundaries around terrestrial and planetary objects can be modified by variations in the upstream solar wind conditions (Zhang et al., 2022; Brain et al., 2017; Yamauchi et al., 2011). Some variations stem in the ion foreshock where backstreaming ions reflected at the bow shock interact with the incident solar wind flow. As such, the foreshock region is subject to a variety of instabilities and transient plasma structures (Collinson et al., 2015; Liu et al., 2020; Schwartz et al., 2018; Sibeck et al., 2021; Fowler et al., 2018; Omidi et al., 2017; Lee et al., 2021).

The solar wind flow also frequently carries discontinuities (rotations), with two common types of rotational (RDs) and tangential discontinuities (TDs) (eugebauer, 2006). RDs are Alfvénic variations and are more frequent than TDs in the solar wind. TDs are non-propagating pressure balanced boundaries that separate two different plasmas that differ in density, temperature, or IMF conditions (Smith, 1973). Unlike TDs, the normal magnetic field component at RDs allows for backstreaming ions or strahl electrons to cross the RD plane (Paschmann et al., 2013). IMF rotations near a bow shock where solar wind interactions are determined by the angle between the shock normal and the upstream interplanetary magnetic field (IMF), $\theta_{Bn}$, become particularly important. Rotations shift the foreshock region by changing $\theta_{Bn}$ across the entire bow shock. Passage of a solar wind flux rope bounded by two discontinuities can create a traveling foreshock (Kajdîc et al., 2017), which is formed when the bundle of field lines between the two solar wind discontinuities connect to the bow shock with quasi-parallel geometry.

Foreshock bubbles (FBs) are ion kinetic scale transient structures formed due to the interaction of backstreaming ions reflected at the shock with a discontinuity in the solar wind plasma (Omidi et al., 2010). As the name suggests they are isolated shock-like structures that develop in the foreshock ahead of the the main bow shock layer. FBs can form across both TDs and RDs (Liu et al., 2015; Wang et al., 2021), though details of the interaction and associated physical processes are remained to be fully understood. Change in the pitch angle and increase in gyrovelocity of backstreaming ions as they cross the discontinuity lead to the concentration of these particles and increase in their perpendicular temperature and thermal energy upstream of the discontinuity (Archer et al., 2015; Turner et al., 2013; Wang et al., 2021). In addition, change in the amplitude and direction of the motional electric field can also concentrate backstreaming ions into a high-pressure plasma layer. These effects consequently lead to the expansion of a plasma bubble against the cold solar wind flow. Unbalanced current densities associated with suprathermal ions, electrons, and solar wind ions create inductive electric fields that drive a magnetic enhancement and a compressional layer on the sunward side of the FB, and a tenuous core on the other side (An et al., 2020; Liu et al., 2020). In spacecraft time series data, FBs are typically manifested as correlated decreases in the solar wind density and the IMF strength which correspond to the core of the FB, followed by a compressional shock layer, the FB shock (Turner et al., 2013). A foreshock compressional boundary may also be present before the FB which exhibits increases in plasma density and magnetic field strength (Rojas-Castillo et al., 2013). The shape of FBs are mostly determined by the orientation of the underlying discontinuity plane, ranging from semi-spherical to planar, and their size scales with the size of the foreshock (Omidi et al., 2020; Turner et al., 2020; Lee et al., 2021). Once formed, the entire FB structure convects with the solar wind.

Due to the small size of the Martian bow shock which is comparable to the solar wind ion convective gyroradius, foreshock ions at Mars play a more significant role in upstream kinetic processes. Foreshock events have direct impact on heating and thermalization of the solar wind, as well as acceleration of charged particles. Both of these effects can have consequences for space weather at Mars which lacks a global dipole magnetic field. In this paper, we present observational evidence for FBs upstream of Mars and discuss the role of the ion dynamics in generating these structures in comparison to their terrestrial counterparts.

## 2 Data Sources and Method

We analyze three events upstream of the Martian bow shock as observed in the Mars Atmosphere and Volatile EvolutioN (MAVEN) spacecraft data (Jakosky et al., 2015). We determine and interpret these events as FBs, noting that there are other transient structures that share some of their properties (Thomsen et al., 1986; Valek et al., 2017). In our analysis, we use magnetic field vectors with 32 Hz sampling rate (Connerney et al., 2015), along with 32 Hz Langmuir probe and waves sweep voltages (LPW-$V_2$) (Andersson et al., 2015), a measure of probe voltage variations corresponding to changes in the surrounding plasma. Electron data from the solar wind electron analyzer (SWEA, (Mitchell et al., 2016)) and the ion measurements from the solar wind ion analyzer (SWIA, (Halekas et al., 2016)) are used. Densities are calculated by integrating the particles distribution function, $\int f(v)d^3v$ over the velocity space. For ions, $f(v)$ is based on SWIA's entire field-of-view (FOV) and energy bins. For electrons it is based on a Maxwell-Boltzmann fit on energy spectra (assuming an isotropic distribution) $f(E) = \frac{2}{\sqrt{\pi}}(k_bT)^{1.5}E^{0.5}\exp\left[-\frac{(E-V_{sc})}{k_bT}\right]$, where $E$ is the electron energy, $T$ is the average temperature, $k_b$ is the Boltzmann constant and $V_{sc}$ is the spacecraft potential. We calculate the ion moments for suprathermal ions from 3D ion distributions when possible, and rely on the difference between SWIA's solar wind only and total ion moments at other times. These data products have different time resolutions from one another and from the magnetic field data (Madanian et al., 2020; Halekas et al., 2017). The local orientation of the FB shock ($\hat{\mathbf{n}}_{FB,shock}$) is determined using the minimum variance analysis (MVA) of the magnetic field across the shock layer. When a plasma sheath layer is present before the FB shock and is resolved in observations, $\hat{\mathbf{n}}_{FB,shock}$ can also be obtained from the mixed coplanarity method for more accurate estimates (Schwartz, 1998). $\theta_{Bn}$ is calculated using normal vectors at the point of IMF intersection with the modeled bow shock surface (Trotignon et al., 2006). The model is a statistical estimate of the shape and standoff distance of the Martian bow shock based on many spacecraft crossings and may not represent the exact conditions of the bow shock layer at a given time. All vector quantities in the paper are in the Mars Solar Orbital (MSO) coordinates in which the *x*-axis is towards the Sun and the *y*-axis points opposite to Mars' orbital motion.

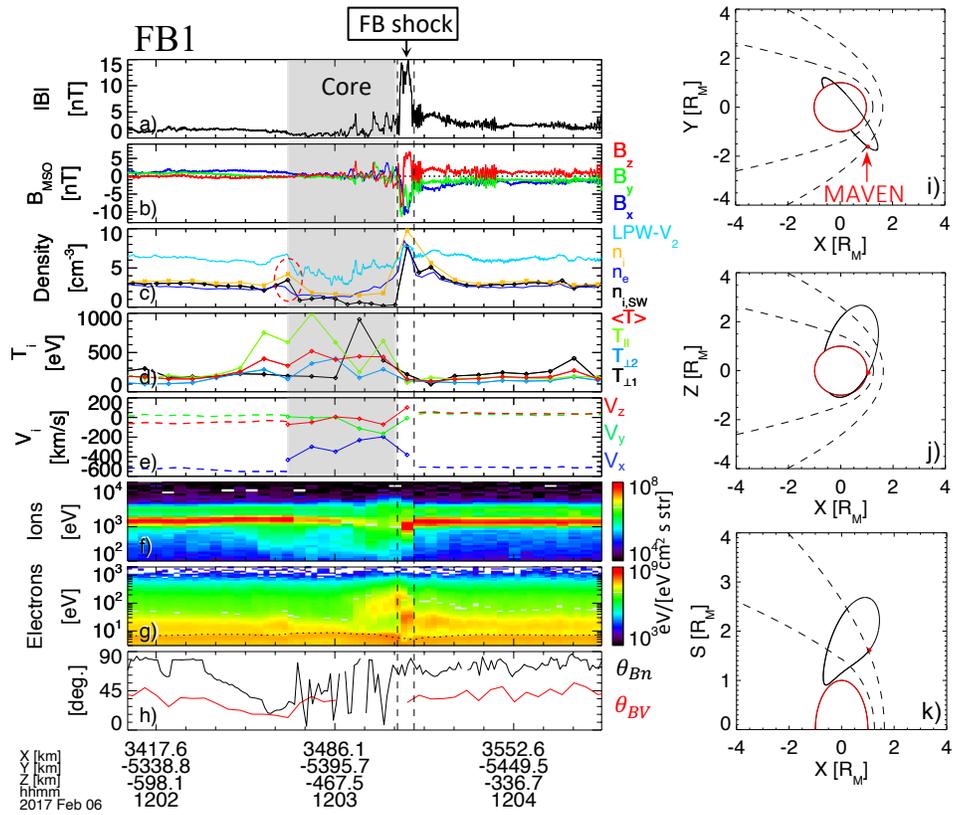

**Figure 1.** Plasma and magnetic field data during Event 1 (FB1) on 6 February 2017 at 12:03:24 UT. Panels show (a) the magnetic field strength, (b) magnetic field components in MSO coordinates, (c) electron (blue), solar wind ion (black,) total ion (yellow) densities, and scaled LPW-V$_2$ voltages (cyan), (d) ion temperature in the local magnetic field frame $T_{\perp 1}$, $T_{\perp 2}$, $T_{\parallel}$, and the average temperature <T>, (e) solar wind (SW, dashed) and non-solar wind (suprathermal) ion (solid) velocities, (f) ion and (g) electron energy flux spectrograms, (h) $\theta_{Bn}$ (black) and $\theta_{BV}$ (red, the angle between **B**$_{IMF}$ and **V**$_{SW}$). Panels (i-k) show the spacecraft position along the trajectory around Mars in $xy$, $xz$, and cylindrical coordinate planes. The FB core is marked with the grey shaded patch and the FB shock layer and sheath is identified by vertical dashed lines. The red dashed circle on panel (c) in this figure marks the density feature associated with the foreshock compressional boundary.

## 3 Observations

Figure 1 shows plasma and magnetic field measurements during the FB1 event observed on 6 February 2017. The core region shaded in grey lasts for 36 s in spacecraft data and exhibits rarefied solar wind plasma and reduced IMF strength. An upstream shock layer is also present showing plasma density compression and heating of ions and electrons. The FB1 core also contains several substructures (i.e., magnetic peaks). Scaled LPW-V$_2$ data shown in cyan in Figure 1c reveal that some substructures are accompanied by density enhancements and are compressional in nature. Such substructures can be reminiscent of transformation or merging of two or more transient foreshock structures (Vu et al., 2022; Omidi et al., 2020). Inside the core, the solar wind signal initially weakens and then almost disappears. There are some overlaps between solar wind ions and suprathermal ions in SWIA's FOV, and the velocities inside the FB1 core represent the bulk ion flow velocity. The parallel ion temperature (green line in panel (d)) starts increasing before the core onset and reaches ~1000 eV inside the core. The perpendicular temperature increases to the same level inside the core. The electron energy spectrogram in Figure 1.g indicates that some electrons are accelerated to ~100 eV inside the core and near the FB shock layer, which could be due to Fermi acceleration of electrons between the FB and the main shock layer (Liu et al., 2017). A comprehensive analysis of the acceleration efficiency and occurrence frequency of this processes upstream of Mars is left for a future study.

Several overlapping field rotations and ultra-low frequency waves downstream of FB1 complicate the selection of a unique magnetic field vector, though all three components of the magnetic field switch sign during this event. FB1 features significantly modulate the underlying discontinuity and its normal vector ($\hat{\mathbf{n}}_D$) cannot be determined through conventional methods (Neugebauer, 2006). However, the solar wind conditions remain rather unchanged before and after FB1 and the underlying IMF discontinuity is likely a RD. Using the magnetic field and ion velocity vectors within the FB1 sheath just prior to the steep drop in |**B**|, together with the upstream flow velocity and IMF vectors averaged between 12:03:58 and 12:04:05 UT, we obtain the shock normal vector $\hat{\mathbf{n}}_{FB,shock}$=(0.7 , -0.4, 0.5) which is within 5° of the MVA estimate. From the conservation of mass flux (Equation 10.29 in Schwartz (1998)), we obtain the FB1 shock speed in the spacecraft frame at around 152.8 kms$^{-1}$ (674.2 kms$^{-1}$ in the solar wind frame), which is much higher than the upstream fast magnetosonic speed of 47 kms$^{-1}$ and supports the growth of a compressional magnetosonic shock. Wave activities immediately after the FB1 shock with circularly polarized waves at ~2 Hz in the spacecraft frame are observed which could be precursor Whistlers emitted by the FB shock. FBs that expand fast enough into the incident solar wind possess a well-developed shock layer which can have its own foreshock region and emit Whistler precursor waves (Liu et al., 2016).

FBs are dynamic structures and the level of wave activity within the core or around the FB shock can vary in different phases of formation. FB2 in Figure 2 is formed at the edge of a moving foreshock bounded by two IMF discontinuities. The magnetic field lines between the two rotations connect to the quasi-parallel bow shock and the event can be viewed as a travelling foreshock (Kajdîc et al., 2017). Nevertheless, the first rotation in IMF at 07:45:43 UT reduces the local $\theta_{Bn}$ from ~64° in the downstream solar wind to ~38°. The flux of suprathermal ions increases inside this transient foreshock.

The FB2 core and shock are formed near the end of the second discontinuity that begins at 07:47:00 UT. The solar wind density is unusually high during this event, $\mathbf{n}_{SW}$~14.9 cm$^{-3}$, but the intensity of the solar wind beam observed at 523 eV significantly reduces inside the core and suprathermal ions make up most of the core plasma density 8.6 cm$^{-3}$. The $\mathbf{n}_{core}/\mathbf{n}_{SW}$ ratio of 0.57 in FB2 is roughly the same as that in FB1 (0.53). The plasma density within FB2's compression layer reaches 66.6 cm$^{-3}$. Theoretically, the spacecraft potential becomes negative for plasma densities ~20 cm$^{-3}$ and as such, electron densities are lower at the shock. Similar to FB1, there are magnetic substructures within the FB2 core. However, LPW-V$_2$ measurements show no sign of commensurate plasma density variations associated with the substructures. It is worth noting that a traveling foreshock in time series data corresponds to a "foreshock cavity" which can harbor magnetic perturbations similar to that observed in FB2's core region (Collinson et al., 2020; Sibeck et al., 2002). In addition, one edge of a foreshock cavity can be associated with a FB (Omidi et al., 2013), as is the case for FB2.

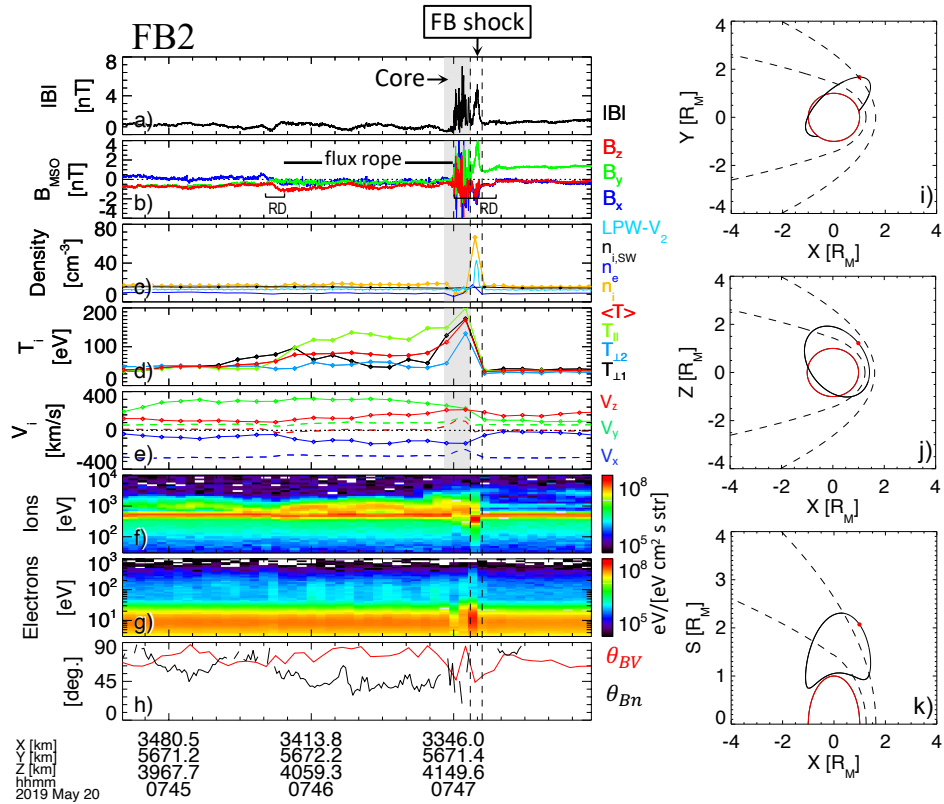

**Figure 2.** An overview of the FB event on 20 May 2019 formed across a travelling foreshock. The figure format is similar to Figure 1. Two RDs associated with the flux rope are indicated in panel b. The y-axis has limited range to emphasize the small amplitude variations within the flux rope.

Simulations of FBs have indicated that when the downstream IMF is roughly parallel to, or makes small angles with the discontinuity plane, as in TDs, the global FB structure is planar in shape. Figure 3 shows an example of a FB forming across a TD. The IMF strength decreases by 0.8 nT from the downstream to the upstream side, while the electron and ion densities each increases by 0.5 cm$^{-3}$. The total solar wind pressure remains constant at 1.02 eV/cm$^{-3}$, consistent with the TD being the sparatrix between two pressure balanced solar wind plasmas. The normal vector to a TD plane is perpendicular to both upstream and downstream magnetic fields, and for the TD in FB3, it is along $\hat{\mathbf{n}}_D$ = (0.4, -0.5, 0.7). The MVA estimate of the local normal to the FB3 shock is $\hat{\mathbf{n}}_{FB,shock}=(0.5, -0.3, 0.8)$, which is within 15° of $\hat{\mathbf{n}}_D$, suggesting that the upstream shock layer is likely formed along the TD plane, and has a planar shape. The FB3 core region corresponds to a period of simultaneous decreases in plasma density and IMF strength and increased flux of suprathermal ions. The shock layer exhibits rather low or modest magnetic field and density compression rates.

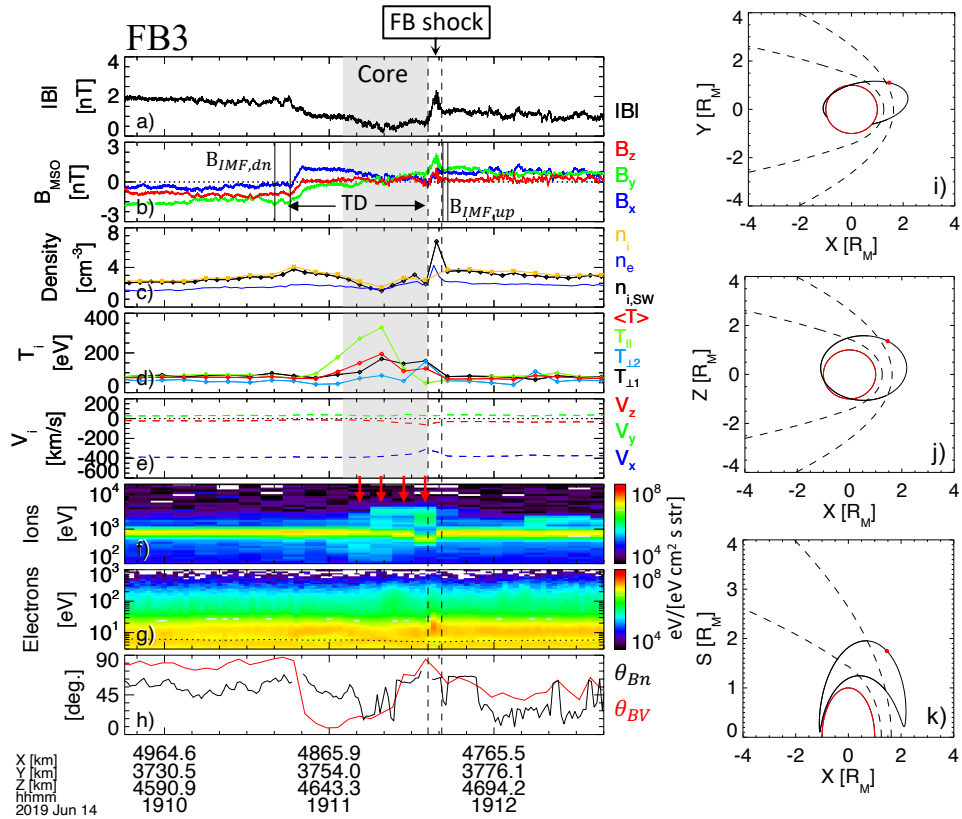

**Figure 3.** The FB3 event observed on 14 June 2019 forming across a TD. The figure format is similar to Figure 4.

Relatively low levels of magnetic perturbations around FB3 compared to FB1 and FB2, and several full 3D ion measurements inside the core makes this event ideal for a more detailed analysis of the dynamics of various ion populations. In Figure 4 we show ion velocity distributions at four timestamps corresponding to red arrows in Figure 3f. The solar wind beam (SW) in these distributions is the intense flux at $V_x \sim -350$ kms$^{-1}$ and is used as a reference to identify the other populations. Backstreaming ions propagate opposite to the solar wind while reflected ions accelerated by the motional electric field have a velocity component perpendicular to it and propagate towards downstream (Madanian et al., 2020). Distributions (a) and (b) indicate that in the first half of the core, as $\theta_{BV}$ and $\theta_{Bn}$ decrease (Figure 2h), the flux of backstreaming ions (BS) increases which results in increased parallel ion temperature. These are reflected ions from the quasi-parallel bow shock and correspond to ion fluxes below the solar wind beam energy in Figure 2f. With increasing $\theta_{Bn}$ and $\theta_{BV}$ in the second half of the core, backstreaming ions disappear from the FOV and the motional electric field drives the reflected ion beam downstream before they reach the spacecraft.

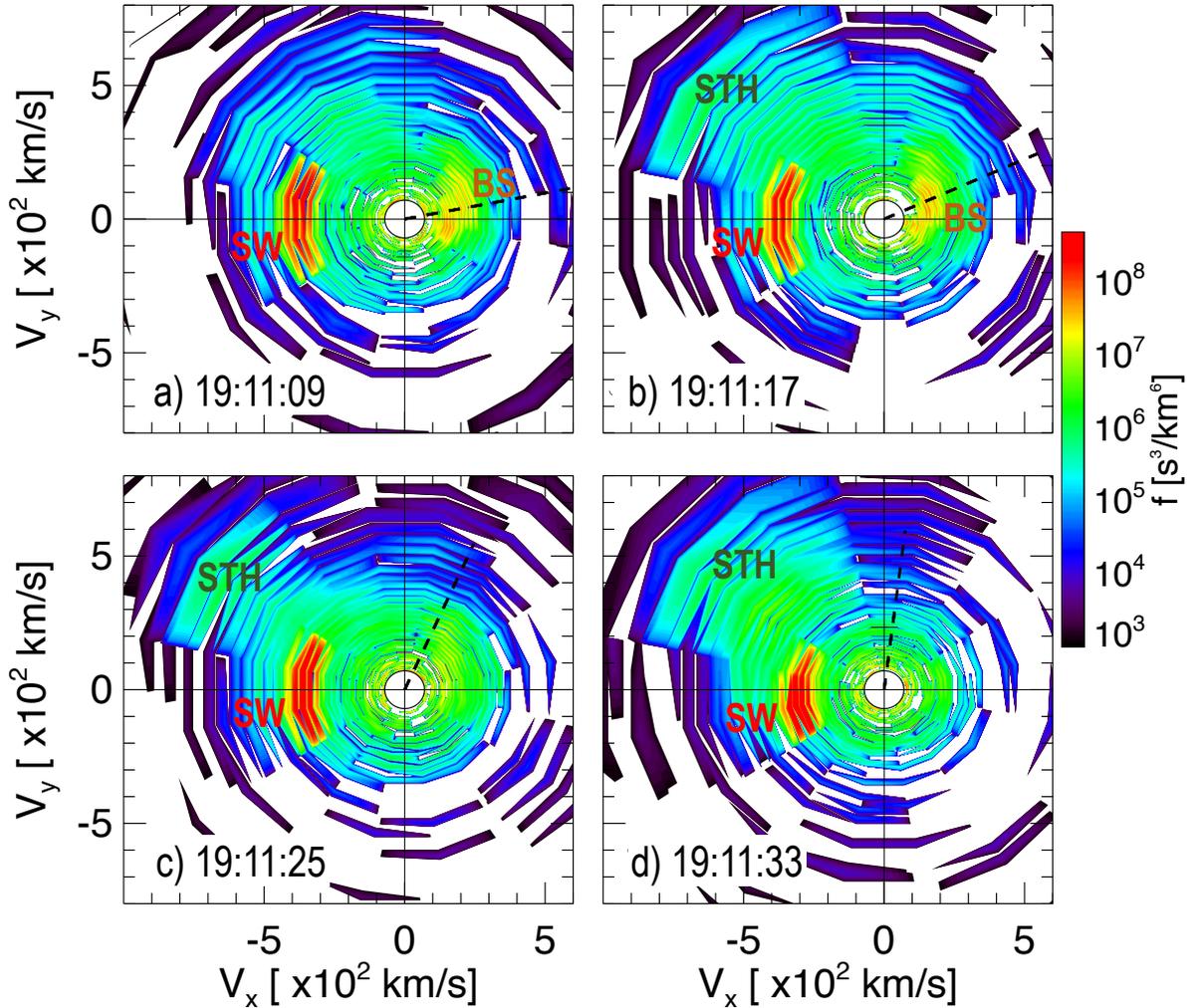

**Figure 4.** 2D cuts through ion phase space velocity distributions at four timestamps across the event on 14 June 2019 (red arrows in Figure 3). The solar wind beam (SW) backstreaming (BS) and accelerated suprathermal (STH) ion populations are labeled on each panel. The dashed black lines show the projection of the magnetic field direction in the plane of the distributions.

As the IMF changes direction (dashed black lines on Figure 4 panels) the suprathermal ions ("STH") continue to enter the instrument from roughly the same direction which indicates that these ions are not sourced locally. These ions propagate anti-sunward and their energies are above the solar wind beam energy, indicating that they have likely been reflected at the quasi-perpendicular side of the bow shock.

**4 Discussions**

Due to the small curvature radius of the Martial bow shock compared to the upstream ion gyroradius, energization of solar wind ions via multiple shock crossing, as in planar shocks, may not be feasible (urgess, 1987; Burgess et al., 2012). However, reflected ion trajectories can overlap between quasi-parallel and quasi-perpendicular regions (Moses et al., 1988). Past hybrid simulations and observational studies at Earth have demonstrated that FBs form when a discontinuity convected by the solar wind encounters backstreaming ions in the foreshock region of the quasi-parallel bow shock (Omidi et al., 2010; Liu et al., 2016). In Figures 1, 2, 3 signatures of ion temperature variations across FBs resembling the observations of FBs at Earth were discussed. To demonstrate the implications of different size scales at Mars and identify the source region of peculiar ion distributions that cause the temperature variations on transient foreshock structures we developed a test particle model to identify the source and trajectory of ion populations in FB3. We track a number of solar wind test particle protons in static and uniform electric and magnetic fields, and update the 3D position and velocity of particles at each step by integrating the Lorentz force on the particles. Each particle undergoes one specular reflection from the nominal bow shock. Wave-particle interactions and other foreshock effects are not included in this simple model.

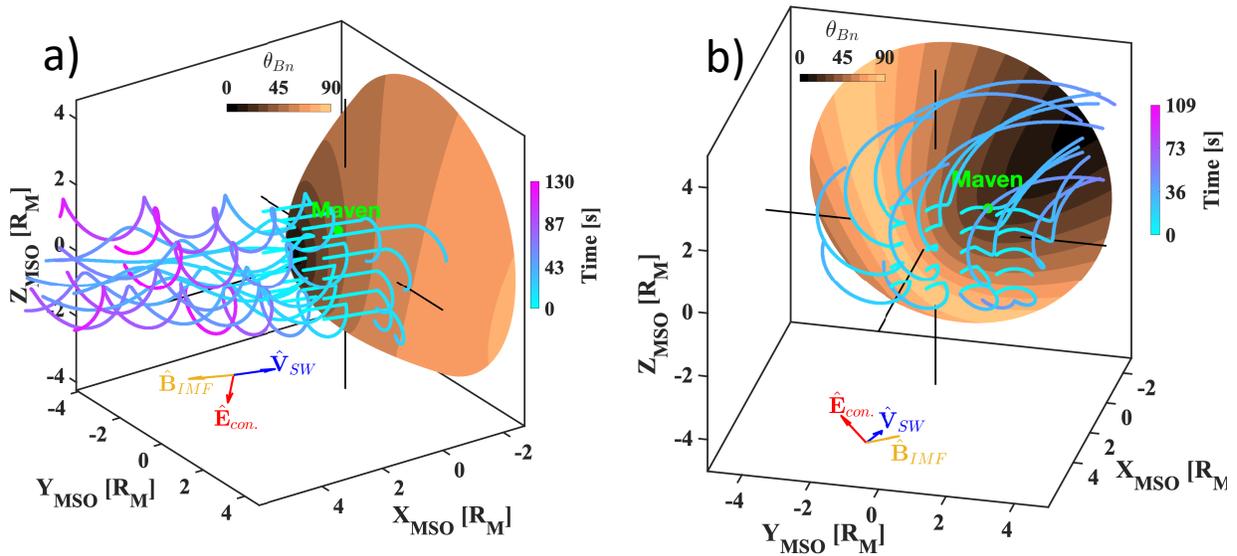

**Figure 5.** 3D view of test particle simulation results showing solar wind proton trajectories in static fields across FB3. Mars's bow shock surface is color coded by $\theta_{Bn}$ (top colorbars), while the color along each trajectory indicates the simulation time (colorbars on the right). The MAVEN spacecraft position is indicated with a green marker. Panels (a) and (b) correspond to the solar wind conditions at the time of distributions (a) and (c) respectively, in Figure 4.

The left panel in Figure 5 shows the ion trajectories under upstream conditions at 19:11:09 UT before the FB3 onset. Backstreaming ions appear near the nose of the bow shock, some of which have trajectories that cross the MAVEN position. These ions are consistent with the "BS" ion population in Figures 4a and b. In all three events, the ion foreshock and associated parallel temperature enhancements appear with the onset of the discontinuity and are caused by the slow IMF rotation. This is a distinguishing aspect of Martian FBs from terrestrial cases which are formed when an upstream discontinuity convected by the solar wind encounters backstreaming ions in the foreshock region of Earth's quasi-parallel bow shock (Omidi et al., 2010; Liu et al., 2016).

Figure 5b shows ion trajectories under the solar wind and IMF conditions in the middle of FB3's core at 19:11:25 UT are shown. Non-solar wind ions that reach MAVEN at this time have been reflected from the quasi-perpendicular regions of the bow shock (Madanian et al., 2020), and after having been accelerated by the motional electric field interact with the upstream discontinuity. These ions form the "STH" population in Figure 4 b, c, and d causing the increase in the perpendicular ion temperature (Figure 3d).

## 5 Conclusions

We report for the first time detailed analysis of FB events upstream of Mars formed across a rotational discontinuity (FB1), a travelling foreshock (FB2), and a tangential discontinuity (FB3). Waves associated with exospheric pickup ions were either absent or weak during these events. For each event we discuss signatures in the magnetic field and plasma moments that are similar and consistent with FB signatures at the terrestrial bow shock. The observations in this study confirm the expectation of FBs being associated with the presence of backstreaming ions and solar wind discontinuities.

The results of our test particle model demonstrate that based on the size of the Martian bow shock and the gyro-radius of the ions reflected from the quasi-perpendicular bow shock it is possible for them to travel sufficiently far upstream to contribute to FB formation at Mars. Even though the Martian foreshock is small compared to that at Earth, backstreaming ion fluxes emerging during slow IMF rotations, and reflected ions from the quasi-perpendicular bow shock that reach the discontinuity can be sufficient to provide the necessary pressure increase for FB formation.

Embedded solar wind discontinuities are a frequent phenomenon at 1 AU (Neugebauer, 2006; Schwartz et al., 2000), and associated foreshock structures have a significant impact on Earth's magnetosphere and ionosphere (Archer et al., 2015; Sibeck et al., 1999; Wang et al., 2018). At Venus, FBs disturb the pressure balance at ionospheric altitudes and uplift the ionospheric heavy ion species, making them more accessible to the pickup process (Omidi et al., 2020). Given the general similarities in solar wind conditions upstream of Mars and Earth (Liu et al., 2021), and the increase in the solar wind speed and Mach number which favor generation of transient foreshock structures, it is likely that these structures cause disturbances at Mars, particularly due to the lack of a global dipole magnetic field. The global impact of these kinetic scale processes on Mars can be better quantified with multi-point observations. At the time of events discussed here, the Mars Express (MEx) spacecraft was in orbit around Mars, however; it was not in a good position to provide complimentary information to our analysis. Future multi-point analyses will benefit from all active spacecraft around Mars.

## Acknowledgments

We acknowledge the MAVEN contract for support. The original funding proposal for this research was not supported by NASA. PK's work was supported by DGAPA/PAPIIT grant IN105620.

## Open Research

All data used in this study are publicly available through the Planetary Data System at https://pds-ppi.igpp.ucla.edu/mission/MAVEN/, under each instrument name, e.g., https://doi.org/10.17189/1414178 for magnetometer and https://doi.org/10.17189/1410658 for LPW data. Further information on each data set is included in references in Section 2.